# Manipulating Biopolymer Dynamics by Anisotropic Nanoconfinement

*Shao-Qing Zhang and Margaret S. Cheung**

Department of Physics, University of Houston, 4800 Calhoun Road, Houston, TX, 77204

szhang9@uh.edu, mscheung@uh.edu

*Corresponding author. Tel.:(713)743-8358, Fax:(713)743-3589, E-mail: mscheung@uh.edu.

**Abstract** How the geometry of nano-sized confinement affects dynamics of biomaterials is interesting yet poorly understood. An elucidation of structural details upon nano-sized confinement may benefit manufacturing pharmaceuticals in biomaterial sciences and medicine. The behavior of biopolymers in nano-sized confinement is investigated using coarse-grained models and molecular simulations. Particularly, we address the effects of shapes of a confinement on protein folding dynamics by measuring folding rates and dissecting structural properties of the transition states in nano-sized spheres and ellipsoids. We find that when the form of a confinement resembles the geometrical properties of the transition states, the rates of folding kinetics are most enhanced. This knowledge of shape selectivity in identifying optimal conditions for reactions will have a broad impact in nanotechnology and pharmaceutical sciences.

**Introduction:** At the interior of a cell, biopolymers such as polypeptides and nucleic acids carry out biological functions in a small space that can be approximated by nano-sized confinement (or nano-sized cavity). For example, proteins fold in chaperonin cages[1,2] and ribosomal exit tunnels[3]; DNA packs in virus capsids[4] and orient in pores[5]. Elucidation of biopolymer dynamics in nano-sized confinement will enhance our understanding of how to manipulate biomaterials using pharmacological chaperones[6,7]. Therefore, we must address the effects of confinement on the dynamics of biopolymers at a molecular detail.

We choose protein folding problem as our working model to investigate confinement effects on biopolymer dynamics because many important issues have been elucidated by experiments[8-17] and theoretical studies[18-29]. For example, in a nanopore (or nanocavity) in which the interior of a pore (or cavity) provides repulsive interaction to proteins, the behavior of a protein in this pore (or cavity) is dramatically altered comparing to its bulk properties. It results in a reduction of available configurational space such that the conformations of unfolded states become more compact. As a consequence, the stability of a native protein is relatively enhanced due to the destabilization of denatured states, assuming that the structure of the native state of a protein remains the same under these confining conditions[18,20,21,23,26,30].

Most of the abovementioned theoretical studies on confinement have been done based on spheres and cylinders [21,23,25,28,30]. These studies motivate us to further inquire the importance of the shape of a confinement on protein folding as shapes of most proteins are ellipsoidal and asymmetric. In this Letter we addressed the anisotropic confinement effects on the thermodynamics and kinetics of protein folding using coarse-grained molecular simulations. Our findings will inspire design of future nanotechnologies for manipulating bionanomaterials.

**Methods:** Streptococcal G B1 domain[31] (PDB ID code 2GB1, see Figure 1(A)) are represented in an off-lattice coarse-grained $C_\alpha$ side-chain model[21,32] (see Figure 1(B)). In this model two beads are represented for each amino acid, except glycine. Chemical bonds and angles are modeled as harmonic springs. Solvent-mediated interactions are coarse-grained into pair-wise Lennard-Jones interactions. To mimic protein-like behavior, a particular set of pairwise interactions for attraction is justified from the native structure (i.e. Gō-like model[33]). Details for this model and Hamiltonian are provided in the Supporting Information (SI). Protein G B1 domain, a 56-residue protein that folds in a two-state fashion, has been well-studied *in* *vitro*[31,34-41], *in silico*[42-46], and *in vivo*[41]. Kinetics and thermodynamics properties of protein folding in both spherical and ellipsoidal confinements (see Figure 1(C) and 1(D)) are obtained by molecular simulations. The numbers of trajectories range from 300 to 500 to ensure our results are statistically significant. Detailed descriptions of molecular simulations are also provided in SI.

**Results:**

*Confinement enhances protein stability*

Shape matters to protein folding at a certain range of volumes of a confinement. Therefore, in the first step we look for a spherical nanopore that will best enhance the stability and folding rates of a protein. Thermodynamics properties of protein folding in various sizes of nanopores are provided in SI. In addition, overall structural fluctuations quantified by the root-mean-square deviation (RMSD) greatly diminished under confinement. The range of these immobilized residues agrees with experimental findings using in-cell NMR spectroscopy[41] (see SI).

Folding temperatures ($T_f$) of protein folding in various sizes of spherical confinement are listed in Table 1. For the bulk case (without confinement), $T_f$ is 360 K, which is the same as the experimental value[31]. When a protein is positioned in a spherical confinement with a considerably large radius (e.g. $R_s = 5R_g^N$, where $R_g^N$ is the radius of gyration of the native state of protein G B1, 1.06 nm, and $R_s$ is the radius of the spherical confinement), $T_f$ remains almost the same as that of bulk, indicating that confinement has little effect on protein stability.

When $R_s < 5R_g^N$, $T_f$ increases inversely with pore sizes, indicating an enhancement of stability of a protein in small confinement. Noticeably, $T_f$ increases by 31 K when a protein is confined in a spherical pore with $R_s = 2.5R_g^N$ and by 70 K with $R_s = 2R_g^N$. An increase of protein stability has been observed in nanoporous sol-gel experiments[9] in which the transition temperature of encapsulated proteins increases by up to 32 K that is about the same range as in our finding and other theoretical studies[28].

*Effects of spherical confinement on folding rates* We next investigate the rates of protein folding in a nanopore. We first aim to pursue a particular pore size that best enhances folding rates and then investigate the effects of shapes on rates of protein folding. Among hundreds of folding simulations, the percentage of unfolded trajectories, $P_u(t)$, as a function of time (t) is provided in Figure 2 for various confining conditions. Folding time is justified by the first passage time of a protein folding event from randomly selected unfolded structures to the native structure at $T_s=343K$. At $T_s$ the free energy profile still remains a energy barrier at a fraction of $k_B T_s$ so that the system avoids the scenario of down-hill folding[47].

$P_u(t)$ is fitted by a bi-exponential function (i.e. $P_u(t) = a*\exp(-bt) + c*\exp(-dt)$) that carries characteristics of kinetic partitioning[48] between a fast-track and a slow-track of folding pathways. We set the maximum folding time at 125μs due to limitation of computing resources. Fitted parameters for different confining conditions are given in Table 2. Here *b* and *d* represent fitted rates of the fast- and slow- tracks of folding pathway, while *a* and *c* correspond to the kinetic partition factors of both tracks.

In the bulk case, kinetic partition factors between the fast-track and the slow-track folding pathways are comparable to each other. When $3R_g^N \leq R_s \leq 5R_g^N$, kinetic partition factors of the slow-track folding pathway decrease with pore sizes. Interestingly, When $R_s=3R_g^N$, folding kinetics is solely dominated by the fast-track folding pathway and $P_u(t)$ becomes almost mono-exponential. Interestingly, the maximum folding rates take place in $R_s=2.5R_g^N$, albeit the slow-track folding pathway starts to reappear. Since the kinetics partition factor of slow-track folding pathway is less than 5%, such a consequence has a minimal effect and the rates are dominated by the fast-track pathway. However, when $R_s=2R_g^N$, due to steric effects from spherical boundary conditions, some of the unfolded structures fail to reach the folded state and that causes long-lived kinetic traps. The observation of an optimal confining size for rate enhancement in protein folding is supported by previous computational studies[21,23,26,30]. Recent experimental measurements[15] verified such predictions by adding or deleting copies of a terminal repeat sequence of GroEL that modulates the volume of a cavity of a GroEL-GroES complex. This studies show that the maximum folding rate of a protein takes place at an intermediate size of a cavity[15].

*Shape of ellipsoidal confinement affects folding rates* Next, we modulate the shape of a sphere of $R_s=2.5R_g^N$. The ratio of the three axes in an ellipsoid is A:B:C and the length of the semi-minor axis is $2.5R_g^N$ (e.g A:B:C of a sphere is 1:1:1). At least one of the three radii of ellipsoidal confinement is $2.5R_g^N$ (2.6 nm). The shape of this confinement varies from a football-like form (e.g. A:B:C=1:1:1.5) to a pancake-like form (e.g. A:B:C=1:1.2:1.2). Rates of protein folding in these various ellipisoids are computed. The fraction of unfolded trajectories, $P_u(t)$, in various shapes of ellipsoidal confinement, at $T_s = 343$ K is given in Figure 3.

Surprisingly, when A:B:C=1:1.2:1.2 (e.g. a pancake), the rates of protein folding are most enhanced (Figure 3). Why can this particular ellipsoidal confinement better promote the rates of protein folding? To answer this question, we dissect the transition state ensemble (TSE) using the progress variable clustering (PVC) method[48-51] and analyze the structural properties of the dominant clusters (Details of PVC methods are provided in SI).

We apply the PVC method to analyze folding trajectories in three confinement conditions, where the ratios of the three axes are 1:1:1 (sphere), 1:1:1.5 (football) and 1:1.2:1.2 (pancake). The structures of the TSE and their dominant clusters are analyzed using the ratios of the three principal radii of gyration (Table 3). Superpositioned structures of transition states of the dominant cluster are illustrated in Figure 4 and they look similar to each other. In addition, the proportion of such a dominant cluster in the TSE is about one third for each condition. It is evident that when the shape of a confinement is similar to that of TSE structures, the rates of protein folding are most enhanced in comparison to other confining environments. *This is the first time the relationship between the folding rate and the shapes of a confinement and a protein has been elucidated.*

Then what makes this particular pancake-like ellipsoid (A:B:C=1:1.2:1.2) a better confining condition while the composition of TSE remains the same? Given the same volume of this ellipsoid, we further modulate its shape of a confinement to the shape of transition states (A:B:C=0.9:1.35:1.18), a sphere (A:B:C=1.13:1.13:1.13), and a football (A:B:C=1:1:1.44). We fit $P_u(t)$ for these three conditions and study how the shape of a confinement affects the kinetics partitioning of folding process, and results are shown in Table 4.

The averaged folding rates in Table 4 are in a range of $0.29\pm0.02\mu s^{-1}$ and $0.34\pm0.02\mu s^{-1}$. These values lie between the averaged folding rates in a sphere of $R_s=3R_g^N$, $0.26\pm0.01\mu s^{-1}$, and that of $R_s=2.5R_g^N$, $0.48\pm0.03\mu s^{-1}$ from Table 2. Together with Table 2 and Table 4 we arrive at a conclusion that the folding rates of the fast track (i.e. *b* in a fitted bi-exponential function for $P_u(t)$) are most determined by the volume.

The shape plays a role in fine-tuning *a, b, c,* and *d,* according to Table 4. It is therefore not surprising that volume has a stronger effect than shape. As long as the shape of a confinement is an oblate (S<0), whether A:B:C=1:1.2:1.2 (S=-0.003) or A:B:C=0.9:1.35:1.18 (S=-0.005) has almost the same confinement effects on folding. Once the confinement becomes a prolate (S=0.037), the folding rates drop by 15%. It is clear that when the shapes of a confinement and a protein resemble to each other, the rates of fast-track folding pathways on the energy landscape[52,53] are most enhanced.

**Discussions and conclusions:** In this Letter we have found the stability of the native state of a protein is enhanced inversely with the size of a confinement at a nano scale. Using profiles of the probability of unfolded trajectories, $P_u(t)$, we are able to justify the effects of the size and the shape of a confinement on trajectories of protein folding. The effects of volume of a confinement on folding rates are most reflected by the dominance of the fast-track folding pathway, while the effects of a shape further fine-tune the folding rates. It is clear now how the rate of protein folding is controlled by the size and the shape of a confinement. Urged by the quest to find the best shape of a confinement that optimizes folding rates the most, we get one important, yet simple, result – *it is when the shape of a confinement resembles that of the transition states of a protein.* We predict that when the shape of transition states are far from a sphere, an enhancement of rates of protein folding due to anisotropic confinement will be most significant.

From mechanistic aspects of protein folding, given the interaction between a protein and the interior of a confinement is repulsive, when the protein and confinement have the same shape, the chance for TSE structures to interact with the surface of a confinement will best enhanced. Therefore an optimally-shaped confinement behaves like a catalyst to accelerate the folding reaction by lowering the free energy barrier of the transition state characterized by a shape-dependent reaction coordinate (Figure 5). Our finding implies an application to manipulating the dynamics of biopolymers by well-designed nano porous materials that use shape as probes. It is also interesting to investigate the kinetics of biopolymers when there are attractive interactions between a protein and the interior of a confinement, which is more complicated[27].

In conclusion, we connect our results with the idea of "shape selectivity"[54,55] from chemistry in which the confining environment is designed for an enhancement of chemical reaction. By studying the structures of transition states *in silico*, we may be able to design pharmaceutical chaperones[6,7] with specific geometry that manipulates protein dynamics in a cell. Our results perhaps can be further applied to modulate interactions between complexes such as like protein-protein association[56], protein-DNA binding[57], and protein-ligand recognition[58], that leads to new directions in bionanotechnology.

**Acknowledgements**: M.S.C. would like to thank Dr. D. Thirumalai for insightful discussions about anisotropic confinement on protein folding. M.S.C. also thanks supports from the University of Houston, TcSUH, and the ACS Petroleum Research Fund (PRF#46762-G2). Computations are partly supported by the Texas Learning Computing Center (TLC$^2$) and an MRAC (TG-MCB070066N) grant from the NSF supported TeraGrid infrastructure. S.-Q. Zhang was partially supported by the Robert A. Welch Foundation (E-1070).

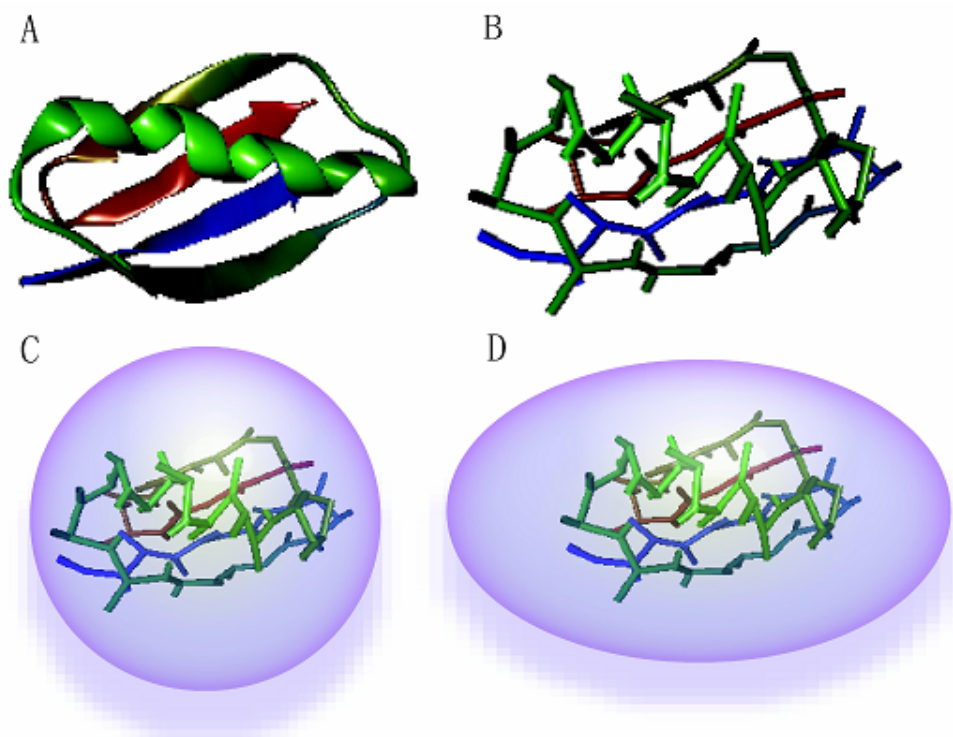

**Figure 1**. The B1 domain of protein G is shown (A) in a cartoon representation of an all-atomistic protein structure (PDBID: 2GB1), (B) in a coarse-grained Cα-Sidechain representation, (C) in a spherical confinement and, (D) in an ellipsoidal confinement.

This figure is created by using VMD[59]. Sizes of a confinement in (C) and (D) are schematic representations and are not to scale with respect to the size of a protein.

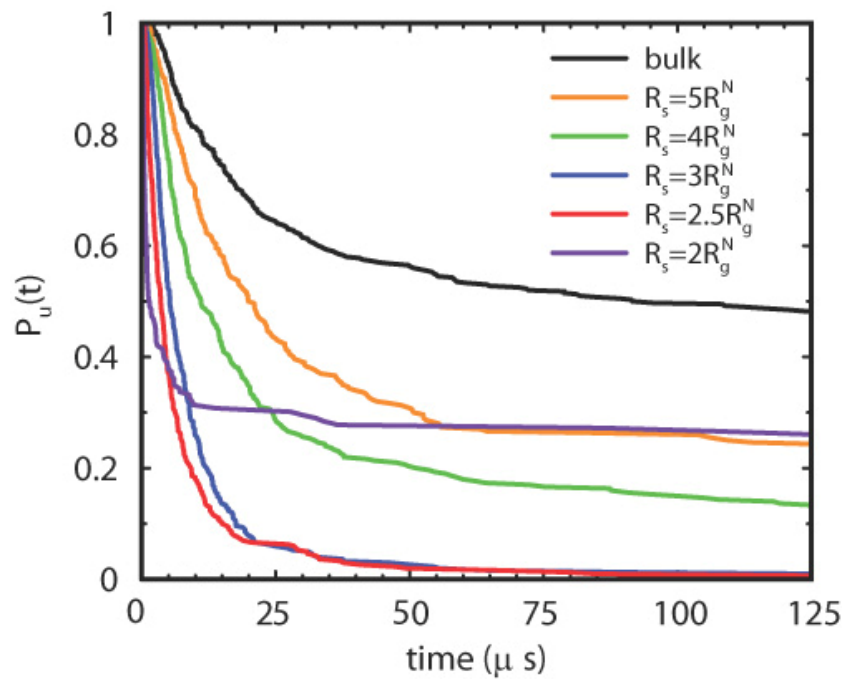

**Figure 2**. Probability of unfolded trajectories, $P_u(t)$, as a function of time in bulk and in various sizes of spherical confinement. The maximum folding time is set to be 125 μs. $R_g^N$ is the radius of gyration of the native state ($R_g^N$=1.06 nm).

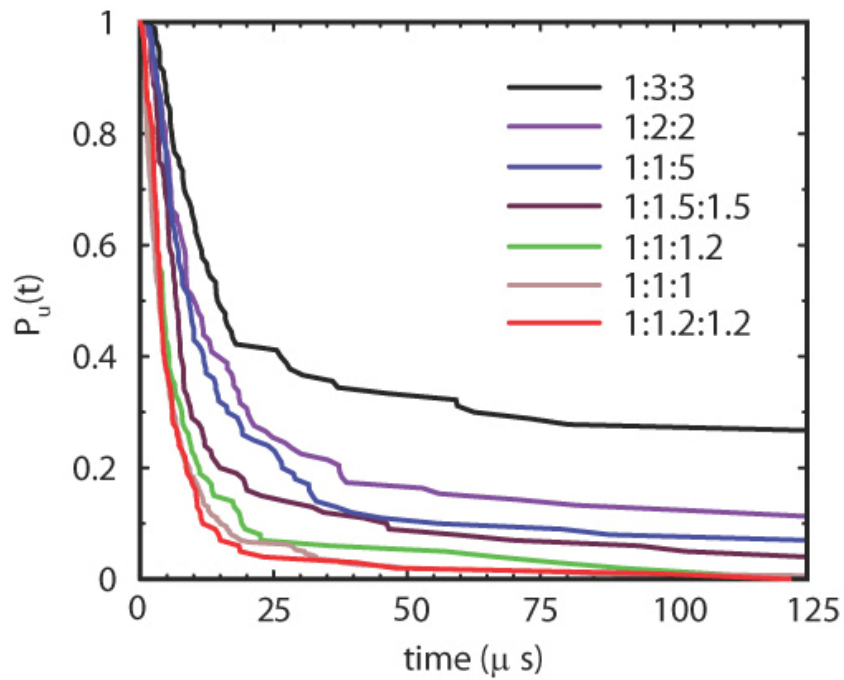

**Figure 3**. Probability of unfolded trajectories, $P_u(t)$, as a function of time in selected various shapes and sizes of ellipsoidal confinement. The ratios represent of the three principle axes of an ellipsoid where the length of the semi-minor axis is $2.5R_g^N$.

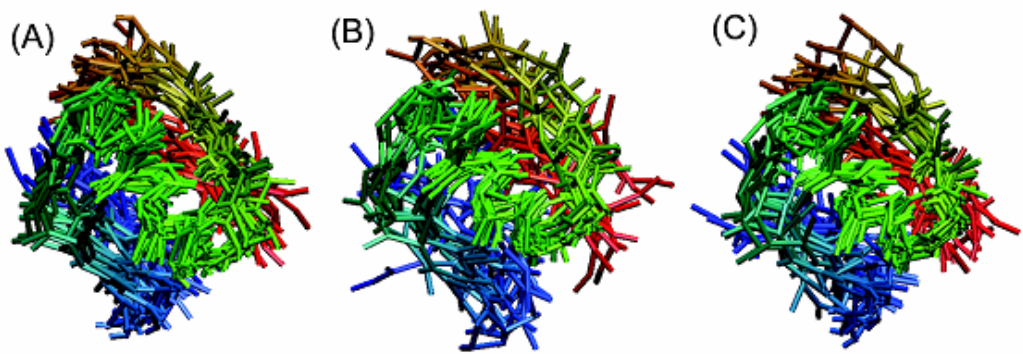

**Figure 4**. The dominant clusters of the transition state ensemble in the confinement at the optimal size $2.5R_g^N$. The ratios of the radii of the ellipsoidal confinement are (A) 1:1:1, (B) 1:1:1.5 and (C) 1:1.2:1.2.

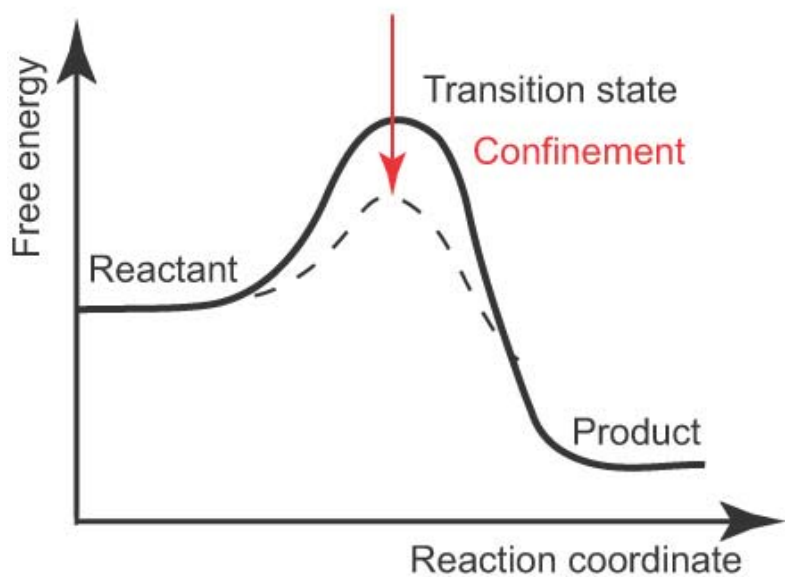

**Figure 5**. An optimally shaped confinement acts as a catalyst in lowering the energy barrier of the transition states along a shape-dependent reaction coordinate.

| Case | bulk | $5R_g^N$ | $4R_g^N$ | $3R_g^N$ | $2.5R_g^N$ | $2R_g^N$ |
|---|---|---|---|---|---|---|
| $T_f$ (K) | 360 | 366 | 372 | 381 | 391 | 430 |

**Table 1**. Folding temperatures of protein G B1 in bulk and in spherical confinement of various sizes.

| Case | a | b (μs$^{-1}$) | c | d (μs$^{-1}$) | rate (μs$^{-1}$) |
|---|---|---|---|---|---|
| Bulk | 0.50 | 4.6×10$^{-2}$ | 0.50 | 8.6×10$^{-5}$ | 0.06±0.00 |
| $5R_g^N$ | 0.77 | 5.0×10$^{-2}$ | 0.23 | 1.8×10$^{-5}$ | 0.10±0.01 |
| $4R_g^N$ | 0.83 | 7.6×10$^{-2}$ | 0.17 | 5.2×10$^{-4}$ | 0.15±0.01 |
| $3R_g^N$ | 0.95 | 1.5×10$^{-1}$ | 0.05 | 1.0×10$^{-2}$ | 0.26±0.01 |
| $2.5R_g^N$ | 0.89 | 2.2×10$^{-1}$ | 0.11 | 2.8×10$^{-2}$ | 0.48±0.03 |
| $2R_g^N$ | 0.63 | 1.2×10$^{-0}$ | 0.37 | 4.2×10$^{-3}$ | 1.71±0.12 |

**Table 2**. Coefficients to fit $P_u(t)$ in Figure 2 using $a*\exp(-b*t)+c*\exp(-d*t)$ and the average folding rates.

| A:B:C | transition state ensemble | the dominant cluster |
|---|---|---|
| 1:1:1 | 1:1.29:1.56 | 1:1.26:1.50 |
| 1:1:1.5 | 1:1.33:1.76 | 1:1.29:1.54 |
| 1:1.2:1.2 | 1:1.31:1.66 | 1:1.28:1.51 |

**Table 3**. Ratios of the three principal radii of gyration of the transition state ensemble and of their corresponding dominant clusters obtained from folding kinetic trajectories in various ellipsoids as shown in Figure 3.

| Case | a | b ($\mu s^{-1}$) | c | d ($\mu s^{-1}$) | rate ($\mu s^{-1}$) |
|---|---|---|---|---|---|
| 1:1.2:1.2, S=-0.003 | 0.94 | $1.7 \times 10^{-1}$ | 0.06 | $9.0 \times 10^{-3}$ | 0.34±0.02 |
| 0.9:1.35:1.18, S=-0.005 | 0.93 | $1.7 \times 10^{-1}$ | 0.07 | $9.0 \times 10^{-3}$ | 0.34±0.02 |
| 1.13:1.13:1.13 S=0 | 0.96 | $1.5 \times 10^{-1}$ | 0.04 | $1.1 \times 10^{-2}$ | 0.34±0.02 |
| 1:1:1.44 S=0.037 | 0.96 | $1.5 \times 10^{-1}$ | 0.04 | $1.2 \times 10^{-2}$ | 0.29±0.02 |

**Table 4**. Coefficients to fit $P_u(t)$ using $a*\exp(-b*t)+c*\exp(-d*t)$ and the average folding rates in different shapes of ellipsoidal confinement of an equal volume $4\pi *1.44(2.5R_g^N)^3/3$.